\documentclass[twocolumn,preprintnumbers,elsart]{revtex4}
\usepackage{makeidx}
\usepackage{amssymb}
\usepackage{amsmath}
\usepackage{mathrsfs}
\usepackage{graphicx}
\usepackage{dcolumn}
\usepackage{bm}
\usepackage[center]{subfigure}
\usepackage{color}
\usepackage{indentfirst}
\usepackage{array}
\usepackage{booktabs}
\usepackage{multirow}

\begin{document}

\title{Two-dimensional anisotropic vortex quantum droplets in dipolar Bose-Einstein
condensates}
\author{Guilong Li$^{1}$}
\author{Xunda Jiang$^{1}$}
\author{Bin Liu$^{1}$}
\author{Zhaopin Chen$^{2}$}
\author{Boris A. Malomed$^{3,4}$}
\author{Yongyao Li$^{1,5}$}
\email{yongyaoli@gmail.com}
\affiliation{$^{1}$School of Physics and Optoelectronic Engineering, Foshan University,
Foshan 528225, China\\
$^{2}$Physics Department and Solid-State Institute, Technion, Haifa 32000,
Israel\\
$^{3}$Department of Physical Electronics, School of Electrical Engineering,
Faculty of Engineering, Tel Aviv University, Tel Aviv 69978, Israel \\
$^{4}$Instituto de Alta Investigaci\'{o}n, Universidad de Tarapac\'{a},
Casilla 7D, Arica, Chile\\
$^{5}$Guangdong-Hong Kong-Macao Joint Laboratory for Intelligent Micro-Nano
Optoelectronic Technology, Foshan University, Foshan 528225, China}

\begin{abstract}
Creation of stable intrinsically anisotropic self-bound states with embedded
vorticity is a challenging issue. Previously, no such states in
Bose-Einstein condensates (BECs) or other physical settings were known.
Dipolar BEC suggests a unique possibility to predict stable anisotropic
vortex quantum droplets (AVQDs). We demonstrate that they can be created
with the vortex' axis oriented \emph{perpendicular} to the polarization of
dipoles. The stability area and characteristics of the AVQDs in the
parameter space are revealed by means of analytical and numerical methods.
Further, the rotation of the polarizing magnetic field is considered, and
the largest angular velocities, up to which spinning AVQDs can follow the
rotation in clockwise and anti-clockwise directions, are found. Collisions
between moving AVQDs are studied too, demonstrating formation of bound
states with a vortex-antivortex-vortex structure. A stability domain for
such stationary bound states is identified. Unstable dipolar states, that
can be readily implemented by means of phase imprinting, quickly transform
into robust AVQDs, which suggests a straightforward possibility for the
creation of these states in the experiment.
\end{abstract}

\maketitle

Nonlocal nonlinearities underlie remarkable phenomena in diverse fields. In
particular, while two and three-dimensional (2D and 3D) self-trapped modes,
supported by the ubiquitous local cubic self-attraction, are unstable due to
the occurrence of the critical and supercritical collapse in the same
settings \cite{1Fibich,2Berge,3Kuznetsov}, the nonlocal nonlinearity arrests
the onset of the collapse, thus stabilizing the 2D and 3D solitons \cite%
{4Malomed,book}. In Bose-Einstein condensates (BECs), stable 2D and 3D
matter-wave solitons were predicted in free space, making use of long-range
van der Waals interactions between Rydberg atoms \cite{8Heidemann,9Maucher}
or laser-induced artificial gravity \cite{Kurizki}, microwave-coupled binary
condensates \cite{16Jieli}, and dipole-dipole interactions (DDIs) \cite%
{11TLahaye,12Pedri,13Nath,14Tikhonenkov,Raghunandan,Blakie}. Unlike other
nonlocal interactions, DDIs feature strong anisotropy in 3D, while in the 2D
geometry DDIs are isotropic or anisotropic if the dipoles are polarized,
respectively, perpendicular to the system's plane or making an angle $<90^{%
\mathrm{o}}$ with it.

The stability of 2D and 3D self-trapped modes with embedded vorticity is a
challenging problem. Solitary vortices are often subject to the azimuthal
modulational instability that develops faster than the collapse, splitting
the vortices into fragments \cite{book}. The nonlocality can help to
suppress the splitting instability if, roughly speaking, the wavelength of
the azimuthal perturbations is smaller than the nonlocality scale. In BEC,
stable vortex solitons supported by nonlocal interactions were predicted for
Rydberg atoms in 3D \cite{15Zhao}, microwave-coupled binary BECs \cite%
{16Jieli}, and dipolar BECs with specially arranged isotropic DDIs in 2D
\cite{17Tikhonenkov}. All these modes featuring isotropic shapes, an open
question is whether anisotropic ones with embedded vorticity (topological
charge) may be made stable in the free space. Anisotropic DDIs offer a
possibility to construct them. In particular, spin-orbit coupling (SOC) has
helped to predict stable anisotropic solitons mixing fundamental
(zero-vorticity) and vortex components in the spinor-dipolar BEC \cite%
{18Xunda,19Bingjin}. However, vortex components play a subordinate role in
the SOC system, carrying a small part of the soliton's norm.

Recently, 3D self-bound states in dipolar BECs were observed in the form of
quantum droplets (QDs), stabilized by the beyond-mean-field (MF) effect,
which is represented by the Lee-Huang-Yang (LHY) term in the respective
Gross-Pitaevskii equation (GPE) \cite{20schmitt,21Chomaz}. The
experimentally demonstrated QDs feature a strong anisotropy in their density
profile in the free space, but they do not carry vorticity. Their
counterparts in the form of isotropic QDs were experimentally created in
quasi-2D \cite{22Cabrea} and 3D \cite{Semeghini,Salasnich} forms in binary
BECs with inter-component attraction, as predicted by Petrov \cite%
{Petrov2015}. The stability and shapes of these self-trapped quantum-fluid
states are determined by the competition between the MF and LHY
nonlinearities \cite{ReviewQD,ReviewQD2,ReviewQD3}.

The LHY-enhanced BEC setting is favorable for stabilizing self-bound vortex
modes. Isotropic 2D and 3D vortex QDs in the free-space binary BEC have been
predicted to be stable with topological charges $S\leq 5$ \cite{23Yongyao}
and $S\leq 2$ \cite{24YVK}, respectively. Stable semi-discrete vortex QDs,
also with $S\leq 5$, were predicted in an array of tunnel-coupled quasi-1D
potential traps \cite{25Xiliang}.

For the dipolar QDs, isotropic vortex solutions with the dipoles polarized
parallel to the vortical pivot were constructed and found to be completely
unstable \cite{26Cidrim}. Hence the inability to create stable vortex modes
in dipolar QDs remains a fundamental and unresolved problem. The aim of this
work is to address this problem, considering dipolar vortex QDs in the 2D
geometry with dipoles polarized parallel to the plane in which the vortex
structure is formed, and the vortex axis being perpendicular to the plane
and the polarization. Compared to previously studies vortex QDs in dipolar
and binary systems, in the current setting their shape is strongly
anisotropic, breaking the rotational symmetry with respect to the vorticity
axis. Previously, no example of stable anisotropic vortex states was
predicted in any model.

Dynamics of the system is governed by the scaled form of the 2D GPE with the
LHY correction:
\begin{equation}
i\frac{\partial }{\partial t}\psi =\left[ -\frac{1}{2}\nabla ^{2}+\Phi _{%
\mathrm{dd}}(\mathbf{r})+g|\psi |^{2}+\gamma |\psi |^{3}\right] \psi ,
\label{GP-LHY}
\end{equation}%
where strengths of the local MF and LHY self-repulsion are, respectively, $%
g>0$ and
\begin{equation}
\gamma =\frac{4g^{5/2}}{3\pi ^{2}}\left( 1+\frac{8\pi ^{2}}{3g^{2}}\right) >0
\label{cLHY}
\end{equation}%
\cite{Daillie2016}. The DDI is represented by term $\Phi _{\mathrm{dd}}(%
\mathbf{r})=\int R(\mathbf{r}-\mathbf{r^{\prime }})|\psi (\mathbf{r^{\prime }%
})|^{2}d\mathbf{r^{\prime }}$, with the interaction kernel
\begin{equation}
R(\mathbf{r}-\mathbf{r^{\prime }})=\frac{1-3\cos ^{2}\Theta }{\left[ b^{2}+(%
\mathbf{r}-\mathbf{r^{\prime }})^{2}\right] ^{3/2}}  \label{kernel}
\end{equation}%
\cite{Sinha,Cuevas}. This kernel implies that, as said above, all dipoles
are polarized along the $x$-direction in the 2D plane, i.e., $\cos
^{2}\Theta =\left( x-x^{\prime }\right) ^{2}/|\mathbf{r}-\mathbf{r^{\prime }}%
|^{2}$, and cutoff scale $b$ is determined by the thickness of the BEC\
layer in the third direction. In this case, the anisotropic DDIs are chiefly
attractive.

Stationary solutions are looked for in the usual form, $\psi (\mathbf{r}%
,t)=\phi (\mathbf{r})e^{-i\mu t}$, with wave function $\phi (\mathbf{r})$
and real chemical potential $\mu $. Dynamical invariants of the system are
the total norm and momentum
\begin{equation*}
N=\int |\phi (\mathbf{r})|^{2}d\mathbf{r},\quad \mathbf{P}=i\int \psi \nabla
\psi ^{\ast }d\mathbf{r},
\end{equation*}%
where $N$ is proportional to the number of atoms in the dipolar BEC, and its
energy,
\begin{equation}
E=\frac{1}{2}\int d\mathbf{r}\left[ |\nabla \psi |^{2}+g|\psi |^{4}+\Phi _{%
\mathrm{dd}}(\mathbf{r})|\psi |^{2}+\frac{4}{5}\gamma |\psi |^{5}\right].
\label{Ham}
\end{equation}

Anisotropic vortex quantum droplets (AVQDs) with integer vorticity $S$ are
produced in the numerical form by means of the imaginary-time method (ITM),
initiated by an anisotropic ansatz,%
\begin{equation}
\phi ^{(0)}(x,y)=A\tilde{r}^{S}\exp \left( -\alpha \tilde{r}^{2}+iS\tilde{%
\theta}\right) ,  \label{ansatz}
\end{equation}%
where $A$ and $\alpha $ are positive constants, and $\left\{ \tilde{r},%
\tilde{\theta}\right\} \equiv \left\{ \sqrt{x^{2}+\beta ^{2}y^{2}},\arctan
(\beta y/x)\right\} $ with an anisotropy factor $\beta >1$. In this work we
set $b=1$ by rescaling and fix $\beta =2$, using $N$ and $g$ as control
parameters. Note that the vorticity may be defined, as per Eq. (\ref{ansatz}%
), in spite of the lack of axial symmetry and nonconservation of the angular
momentum.
\begin{figure}[h]
{\includegraphics[width=0.99\columnwidth]{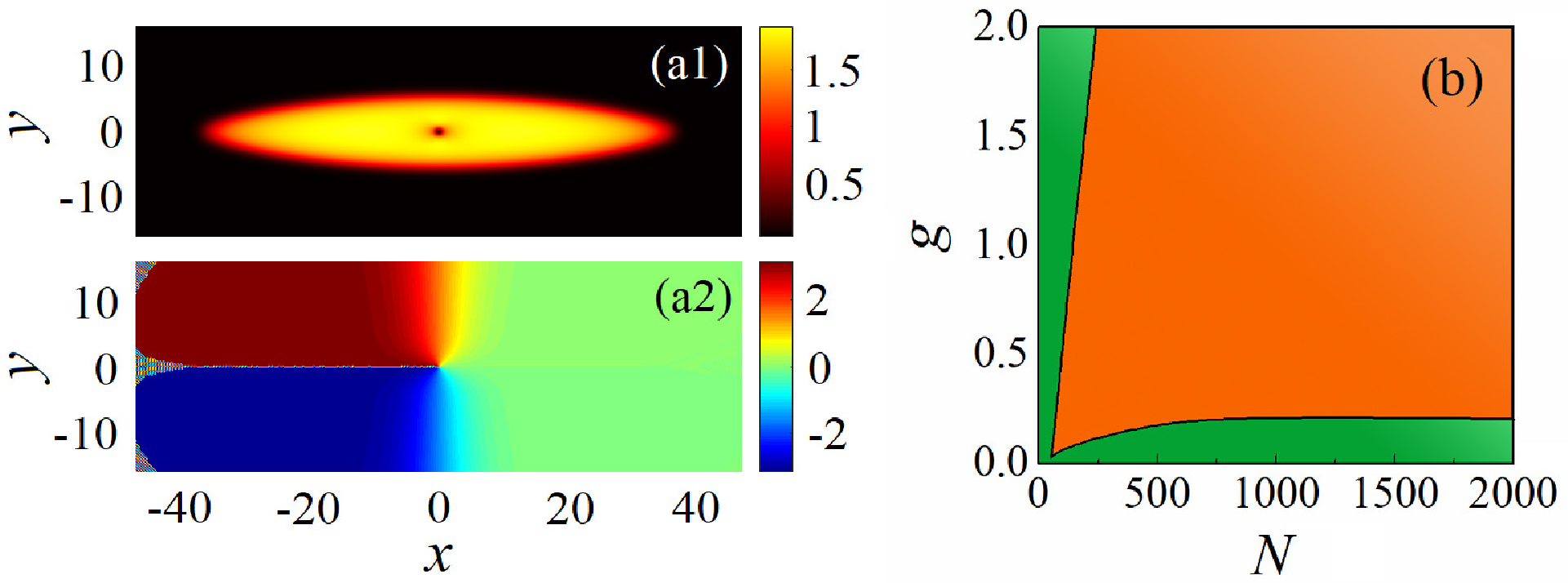}}
\caption{(a1,a2) A typical example of stable AVQDs (anisotropic vortex
quantum droplets),\ with $(N,g)=(1000,0.25)$. The panels (a1,a2) display,
severally, density and phase patterns of the droplet. A movie of the
perturbed evolution of this state is presented in Fig. I of the Supplemental
material. (b) In the plane of $\left( N,g\right) $, AVQDs ($S=1$),
fundamental QDs ($S=0$) are stable in the orange area. In the green area,
only dipole mode and fundamental QDs can be stable.}
\label{Example}
\end{figure}

A typical example of numerically found stable AVQDs with $S=1$ is produced
in Figs. \ref{Example}(a1,a2). Their stability was tested by direct
simulations of the perturbed evolution for a sufficiently long time \cite%
{Distance}. Results of the analysis are summarized in the form of the
stability area in the $(N,g)$ plane, plotted in Fig. \ref{Example}(b). In
particular, the stable AVQDs are found at $N>N_{\min }$, where $N_{\min }$
is a gradually increasing function of $g$. At $N<N_{\min }$, the vortex
ansatz produces stable dipole modes. For fixed $N$, stable AVQDs are found
at $g>g_{\min }(N)$, e.g., $g_{\min }(N=500)\approx 0.2$. At values of $g$
slightly smaller than $g_{\min }$, the AVQD starts spontaneous drift,
keeping its topological charge. Deeper into the region of $g<g_{\min }$, the
unstable AVQDs quickly decay to the fundamental QDs. The dipole modes
produced by the dipole ansatz in the orange region of Fig. \ref{Example}(b)
are unstable. In the course of the perturbed real-time evolution, they
spontaneously transform into stable vortices, suggesting a feasible scenario
for the creation of the vortex modes in the experiment, as discussed in
detail below.

The instability of 3D isotropic vortices, with dipoles polarized parallel to the vorticity axis\cite{26Cidrim}, is explained by the fact the void around the long axis implies effective removal of a tube filled by dipoles which chiefly attract each other, i.e., removal of a large amount of the negative interaction energy (in other words, addition of a positive energy, which naturally leads to the destabilization). On the contrary to that, the present configuration, with the mutually perpendicular vorticity and in-plane polarization, stability is feasible because the void occupies an area around the vortex’ pivot, where the DDIs overall display repulsion, thus removing some positive energy.

\begin{figure}[h]
{\includegraphics[width=0.9\columnwidth]{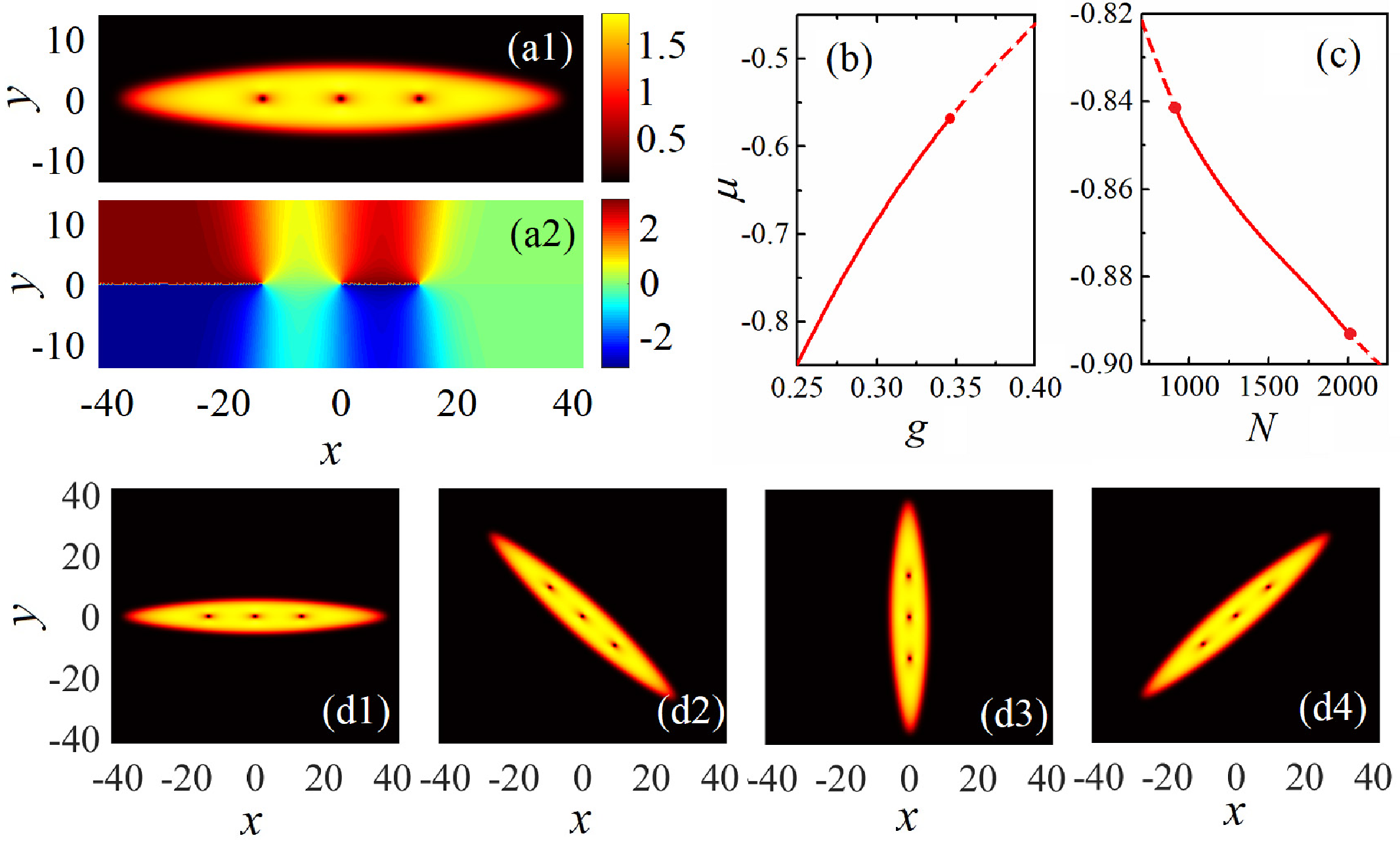}}
\caption{(a) Density and phase patterns of the stable
vortex-antivortex-vortex bound state with $(N,g)=(1000,0.25)$, $x_{0}=13.6$
and $\protect\mu =-0.8482$. A movie of the perturbed evolution of this state
is displayed as Fig. V of Supplemental Material. (b,c) The chemical
potential of the states of this type versus $g$ (at $N=1000$) and $N$ (at $%
g=0.25$), respectively. Solid and dashed parts of the curves represent,
severally, stable and unstable states. (d1-d4) Steady spinning of a
vortex-antivortex-vortex bound state with $(N,g)=(1000,0.25)$, under the
action of the polarizing magnetic field rotating with angular velocity $%
\protect\omega =0.2\protect\pi \times 10^{-3}$. The shape of the bound state
is displayed at $t=0$ (d1), $1250$ (d2) $2500$ (d3), $3750$ (d4). }
\label{v-av-v}
\end{figure}

The stability was tested by real-time evolution with the perturbed noises. The time should exceed the respective diffraction time,
at least, by an order of magnitude, be larger than $10\times 2R^{2}$, where $%
R$ is the radial size of the droplets. For example, in Figs. \ref{Example}%
(a) and \ref{v-av-v}(a), the radial sizes of the 2D-AVQDs and the
vortex-antivortex-vortex self-bound state are $R\sim 30$ and $40$,
respectively, while the corresponding simulation times are $\simeq 18000$
and $32000$. The stability and instability of the 2D-AVQDs and dipole states, identified by
means of direct simulations, was corroborated by numerical computation of
eigenvalues for small perturbations around the stationary states, governed
by the linearized Bogoliubov -- de Gennes (BdG) equations. For the stable
2D-AVQDs, the BdG solution reveals the existence of an intrinsic perturbation
eigenmode. In particular, for the state shown in Figs. \ref{Example}(a1,a2),
the eigenfrequency of the internal mode is $0.5$. Detailed analysis the
respective excited states of 2D-AVQDs will be reported elsewhere.

For stationary QDs with a large norm, one can apply the analytical
Thomas-Fermi (TF) approximation, neglecting the kinetic-energy term in Eq. (%
\ref{GP-LHY}). Then, the QD density, $n$, determines the total energy (\ref%
{Ham}) as
\begin{equation}
E=\frac{1}{2}\left[ (\varepsilon+g)n^{2}+\frac{4}{5}\gamma n_{e}^{5/2}\right]
A,  \label{E}
\end{equation}%
where $A=N/n$ is the area of the QDs, and $\varepsilon =\int d\mathbf{r}R(%
\mathbf{r})\approx -3.23$ represents the nonlocality effect. The equilibrium
density, $n_{e}$, is the value providing an energy minimum as $dE/dn=0$,
which yields
\begin{equation}
\sqrt{n_{e}}=-\frac{5}{6\gamma }(\varepsilon +g),\quad A_{e}=\frac{36}{25}%
\gamma ^{2}\frac{N}{(\varepsilon +g)^{2}}.  \label{ne}
\end{equation}%
An obvious condition, $\sqrt{n_{e}}>0$, applied to Eq. (\ref{ne}), leads to $%
g<g_{\max }\equiv -\varepsilon \approx 3.23$. At $g>g_{\max }$, the strong
local repulsion overcomes the effective nonlocal attraction, hence no
self-bound state can be formed. Actually, $n_{e}$ becomes very small and the
size of AVQD very large at $g>2$, which makes it difficult to reach $g_{\max
}$ in the numerical solution. The chemical potential corresponding to
density (\ref{ne}) is
\begin{equation}
\mu _{e}=(\varepsilon +g)n_{e}+\gamma n_{e}^{3/2}.  \label{mu}
\end{equation}%
These analytical predictions are compared to numerical findings in Figs. \ref%
{char}(a1-a3,b1-b3).

\begin{figure}[h]
{\includegraphics[width=0.99\columnwidth]{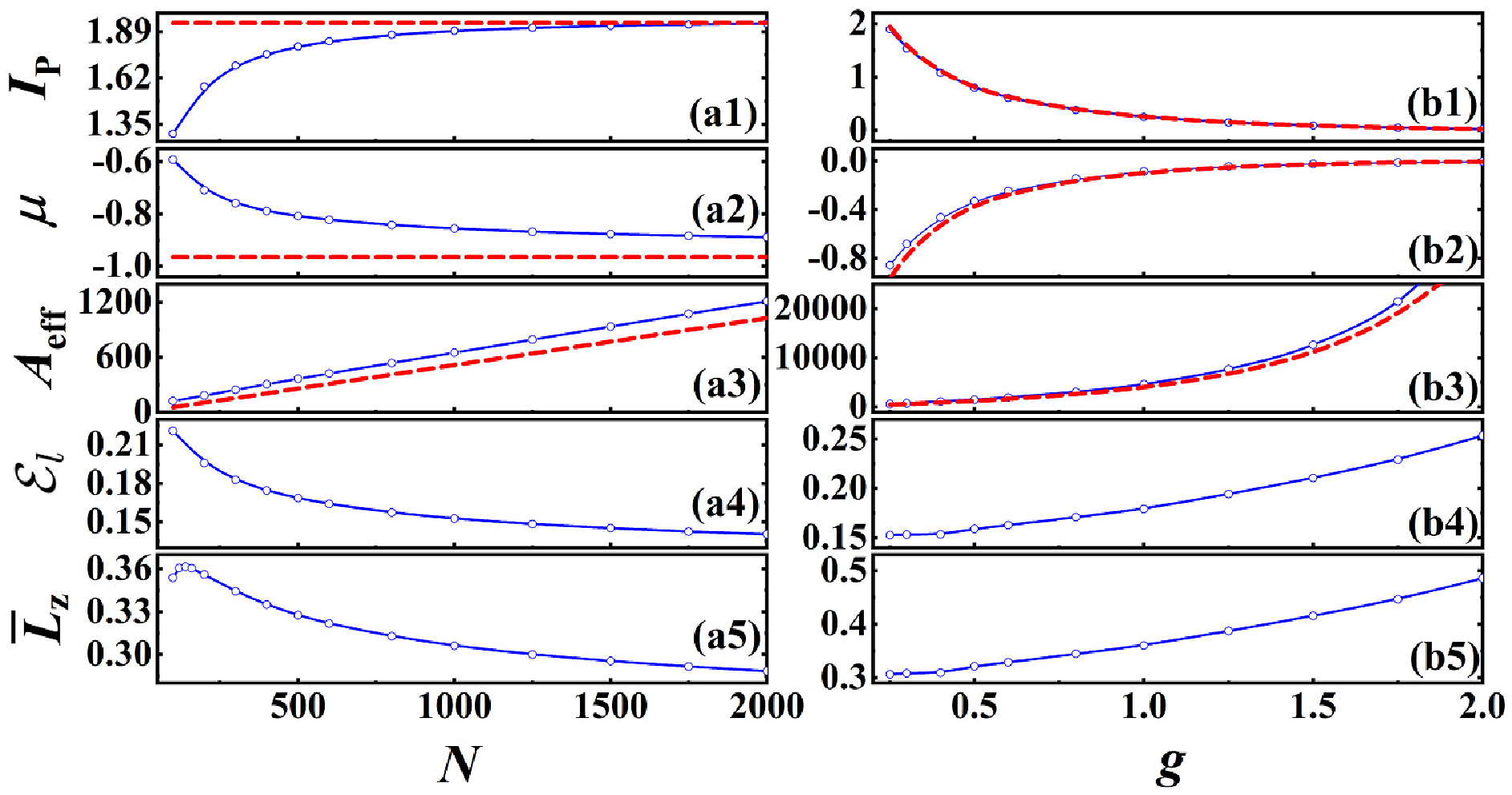}}
\caption{The peak density ($I_{\mathrm{P}}$), chemical potential ($\protect%
\mu $), effective area ($A_{\mathrm{eff}}$), aspect ratio ($\mathcal{E}_{l}$%
), and total orbital momentum ($\bar{L}_{z}$), see Eq. (\protect\ref%
{definition}), vs. $N$ (a1-a5) and $g$ (b1-b5). In panels (a1-a5), $g=0.25$
is fixed, while in panels (b1-b5) $N=1000$. Red dashed curves in panels
(a1-a3,b1-b3) represent the analytical approximation given by Eqs. (\protect
\ref{ne}) and (\protect\ref{mu}), respectively.}
\label{char}
\end{figure}

To characterized AVQD families, we define their effective area, aspect
ratio, and angular momentum:
\begin{equation}
A_{\mathrm{eff}}=\frac{\left( \int |\phi |^{2}d\mathbf{r}\right) ^{2}}{\int
|\phi |^{4}d\mathbf{r}},\mathcal{E}_{l}=\frac{W_{y}}{W_{x}},\bar{L}_{z}=\int
\frac{\phi ^{\ast }\hat{L}_{z}\phi }{N}d\mathbf{r},  \label{definition}
\end{equation}%
where $W_{y}\equiv \left( \int |\phi (x=0,y)|^{2}dy\right) ^{2}/\int |\phi
(x=0,y)|^{4}dy$, $W_{x}\equiv \left( \int |\phi (x,y=0)|^{2}dx\right)
^{2}/\int |\phi (x,y=0)|^{4}dx$, and $\hat{L}_{z}=-i(x\partial
_{y}-y\partial _{x})$. Figure \ref{char} displays these quantities, along
with $I_{\mathrm{P}}=|\phi |_{\max }^{2}$ and $\mu $, versus $N$ and $g$ in
the stability area. In panel \ref{char}(a1), the peak value saturates at $%
\left( I_{\mathrm{P}}\right) _{\mathrm{sat}}\approx 1.938$ if $N$ is
sufficiently large, as expected for an incompressible quantum fluid.
According to Eq. (\ref{ne}), the analytical prediction of the equilibrium
density is $n_{e}\approx 1.942$, in close agreement with $\left( I_{\mathrm{P%
}}\right) _{\mathrm{sat}}$. Panel \ref{char}(a2) shows that the chemical
potential satisfies the Vakhitov-Kolokolov (VK) criterion, $d\mu /dN<0$,
which is the well-known necessary stability condition for self-trapped modes
\cite{VK,1Fibich,2Berge,3Kuznetsov}. For large $N$ the chemical potential
saturates at $\mu \approx -0.889$. The analytical prediction given by Eq. (%
\ref{mu}) is $\mu _{e}\approx -0.965$, which is also relative close to the
numerical finding. In panels \ref{char}(b1,b2), both $I_{P}$ and $\mu $
decay to zero at $g\rightarrow 2$, which agrees well with the analytical
predictions provided by Eqs. (\ref{ne}, \ref{mu}). In panels \ref{char}%
(a3,b3) the effective area closely matches the analytical result (\ref{ne}).
In panels \ref{char}(a4,b4) the aspect ratio remains smaller than $0.25$,
which indicates that AVQDs manifest strong anisotropy with the elongation
along the $x$-direction. For very strong anisotropy, \textit{viz}., at $%
\mathcal{E}_{l}\leq 0.2$, relation $\bar{L}_{z}\approx 2\mathcal{E}_{l}$
roughly holds, which is explained by a straightforward estimate, $y\partial
_{x}\sim W_{y}/W_{x}$, which is valid in this case.


Stationary AVQDs with higher vorticities, $S\geq 2$, have also been found in
the numerical form. However, simulations demonstrate that they are fully
unstable \cite{unstableQD}.

\begin{figure}[h]
{\includegraphics[width=0.99\columnwidth]{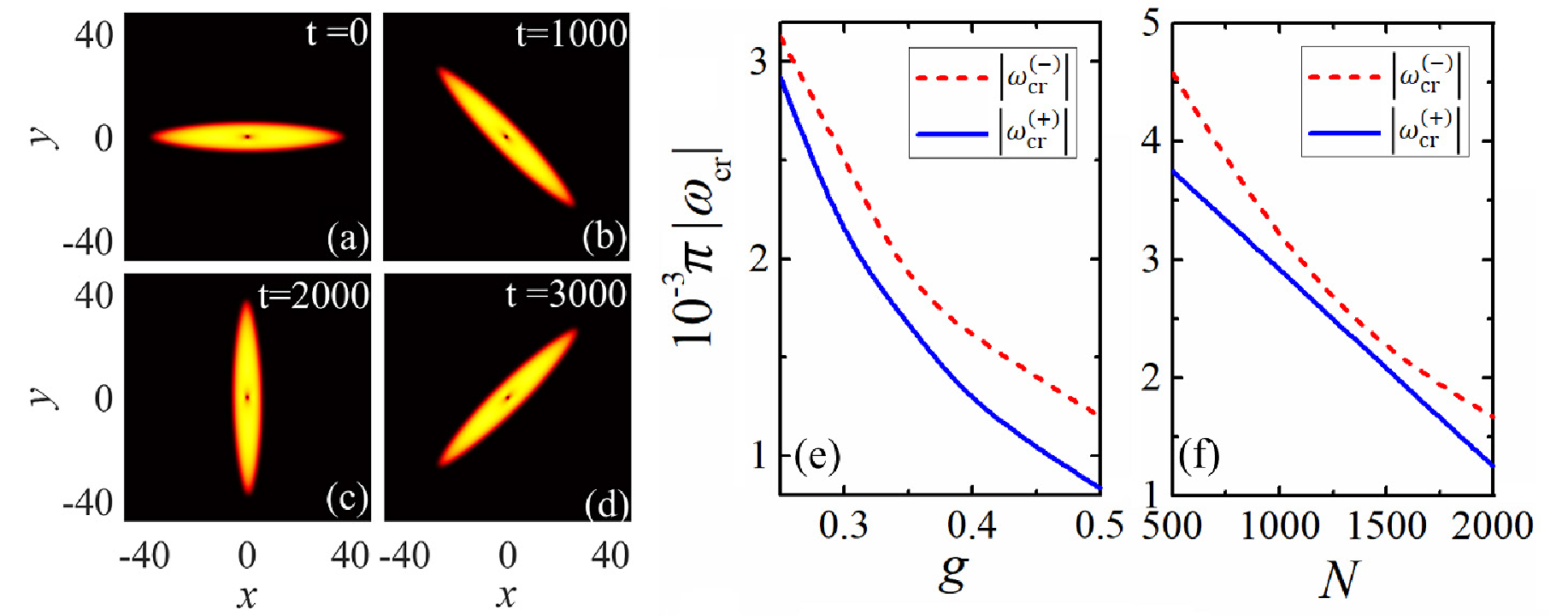}}
\caption{(a-d) Steady spinning of AVQD with $(N,g)=(1000,0.25)$, which
follows the rotation of the polarizing magnetic field with angular velocity $%
\protect\omega =0.25\protect\pi \times 10^{-3}$. The shape of AVQD is
displayed at $t=0$ (a), $1000$ (b) $2000$ (c), $3000$ (d). Simulations of
the stable perturbed rotation of the AVQD are displayed in a movie presented
as Fig. II of Supplemental Material. Panels (e) and (f): the largest angular
velocities, $\left\vert \protect\omega _{\mathrm{cr}}^{\left( \pm \right)
}\right\vert $, which admit stable spinning of AVQD in two opposite
directions, vs. $g$ (at $N=1000$) and $N$ (at $g=0.25$), respectively. }
\label{Rotate}
\end{figure}

It is known that zero-vorticity solitons in the dipolar BECs can rotate,
adiabatically following slow in-plane rotation of the magnetic field which
polarizes atomic magnetic moments \cite{14Tikhonenkov}.\ The rotation can be
introduced in Eq. (\ref{kernel}) by replacing $\Theta \rightarrow \Theta
+\omega t$. When the rotation is sufficiently slow, \textit{viz}., $\omega
<\omega _{\mathrm{cr}}$, AVQDs are able to follow it, in the state of
spinning motion. Figures \ref{Rotate}(a-d) show an example of the steady
rotation of AVQD. Numerical simulations demonstrate that $\omega _{\mathrm{cr%
}}$ decreases with the increase of the QD's size. Indeed, the large size
makes it difficult to synchronize the rotational motion of the QD's core
area and remote edges. The intrinsic vorticity of AVQD makes it more
tolerant to the rotation in the direction of the inner vorticity than
against it, therefore the simulations demonstrate two different critical
values, $\omega _{\mathrm{cr}}^{\left( \pm \right) }$, in Figs. \ref{Rotate}%
(e,f).

\begin{figure}[h]
{\includegraphics[width=1.0\columnwidth]{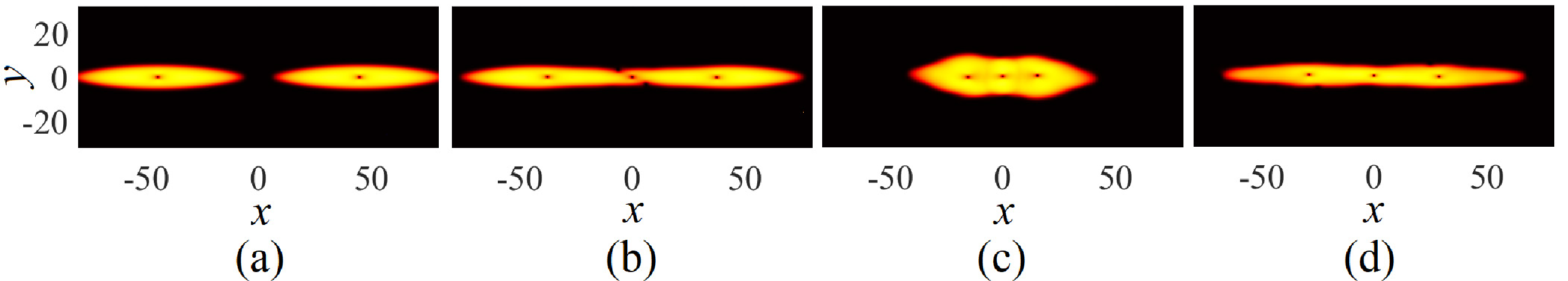}}
\caption{The collision of two AVQDs with identical vorticities, initiated,
at $t=0$, by input $\protect\phi (x-x_{0},y)e^{-i\protect\eta x}+\protect%
\phi (x+x_{0},y)e^{i\protect\eta x}$ with $x_{0}=64$, $\protect\eta =0.025$,
$g=0.25$, and norm of each AVQD $N=1000$. (a-d) Density patterns at $t=550$
(a), $635$ (b), $760$ (c), and $900$ (d). A movie with simulations of this
inelastic collision is presented in Fig. IV of the Supplemental Material.}
\label{collision}
\end{figure}

Stable AVQDs can be set in motion by opposite kicks $\pm \eta $ applied
along the $x$ or $y$-direction. Accordingly, it is possible to simulate
collisions between AVQDs moving in opposite directions. Results demonstrate
a drastic difference from the usual scenario of collisions in non-integrable
systems with local nonlinearity, where the increase of $\eta $ leads to a
transition from inelastic collisions between slow solitons to quasi-elastic
outcomes for fast ones \cite{RMP}. In the present setting, elastic
collisions are only observed between AVQDs moving in the $y$-direction if
kick $\eta $ is relatively small. If $\eta $ is larger, or the head-on
collision happens in other directions in the $(x,y)$ plane, the outcome is
inelastic, leading to merger of AVQDs with identical ($S_{1}=S_{2}=1$) or
opposite ($S_{1}=-S_{2}=1$) vorticities into localized breathing modes. A
noteworthy result is produced by the collision between AVQDs with $%
S_{1}=S_{2}=1$ traveling in the $x$-direction (in which the dipoles are
polarized): formation of a transient state in the form of a breathing
vortex-antivortex-vortex structure. Eventually, this long-lived state
transforms into a zero-vorticity breathing one. A typical example of such a
collision is displayed in Fig. \ref{collision}, where the central pivot,
which represents the antivortex, emerges in the beginning of the merger of
the two colliding vortices [see panels (b) in the figure].

\

The production of the above-mentioned long-lived vortex-antivortex-vortex
breather by collisions along the $x$ direction suggests that the system may
support truly stationary bound states with a similar structure. They can
indeed be produced by means of ITM, starting from a generalization of ansatz
(\ref{ansatz}), 
\begin{equation}
\phi ^{(0)}=\sum_{\pm}A_{\pm}\tilde{r}_{\pm}\cdot e^{-\alpha _{\pm}\tilde{r}%
_{\pm}^{2}+i\tilde{\theta}_{\pm}}+A\tilde{r}e^{-\alpha\tilde{r}^{2}-i\tilde{%
\theta}},  \label{3pivotansatz}
\end{equation}%
where $\tilde{r}_{\pm }\!\equiv\!\sqrt{(x\!\pm\! x_{0})^{2}\!+\!\beta
^{2}y^{2}}$, $\tilde{\theta}_{\pm }\!\equiv\! \arctan \left[ \beta
y/(x\!\pm\! x_{0})\right] $, and $x_{0}$ is an appropriately chosen
separation. A typical example of such a stable bound state is displayed in
Fig. \ref{v-av-v}(a). The family of stable bound states is characterized by
dependencies of the chemical potential on $g$ and $N$, as shown in Figs. \ref%
{v-av-v}(b,c). In the latter panels, stable bound state of the
vortex-antivortex-vortex type populate areas $g<0.35$ and $900<N<2000$. Note
also that the $\mu (N)$ relation satisfies the aforementioned VK criterion, $%
d\mu /dN<0$. Similar to what is presented in Fig. \ref{Rotate}, it is
possible to apply a rotating magnetic field to the bound states of the
present type, and test a possibility of their steady spinning motion, see
Figs. \ref{v-av-v}(d1-d4) and a movie in Fig. VI of Supplemental Material.

\begin{figure}[h]
{\includegraphics[width=0.9\columnwidth]{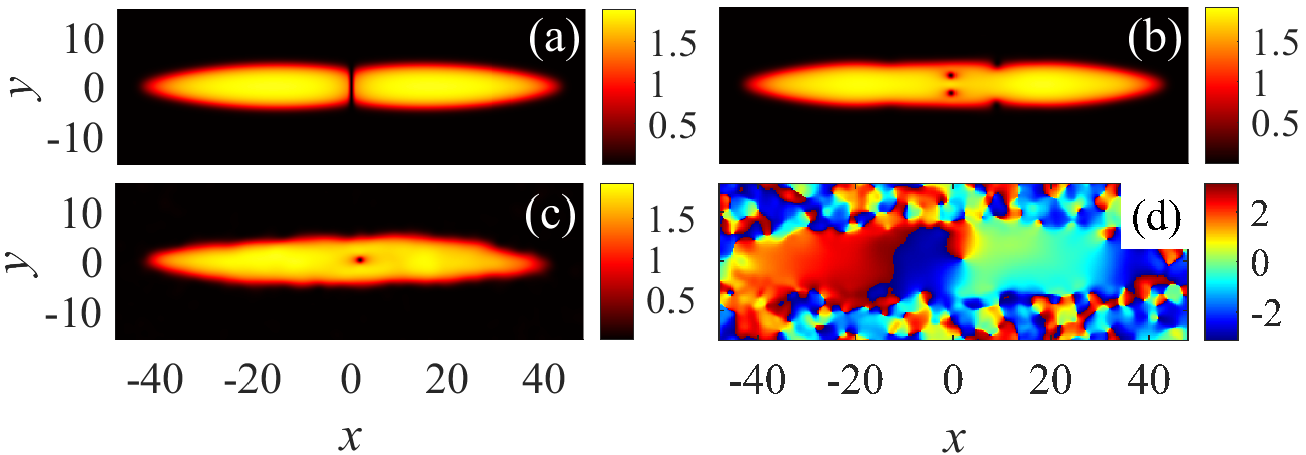}}
\caption{(a-c) Spontaneous transformation of the unstable dipole mode with $%
N=1000$ and $g=0.25$ into a stable AVQD is represented by density snapshots taken at 
$t=0$ (a), $50$ (b), and $700$ (c). (d) The phase pattern at $t=700$. A movie of the evolution of this
dipole mode is displayed as Fig. VII of Supplemental Material.}
\label{Dipolemode}
\end{figure}

As mentioned above, dipole modes coexisting with the stable AVQDs in the
chart displayed in Fig. \ref{char}(b) are unstable, rapidly transforming
into stable vortex modes in the course of the perturbed real-time evolution,
see an example in Fig. \ref{Dipolemode}. This dynamical process suggests a
promising way to create AVQDs in the experiment. Indeed, one can first
produce an unstable dipole mode by imprinting a phase kink onto the
fundamental QD, to be followed by its spontaneous transformation into the
AVQD. The phase-imprinting method is widely available, being used, in
particular, to create dark solitons in BEC \cite{Burger}. Spontaneous decay
of quasi-1D dark solitons into delocalized vortices has been reported too
\cite{Anderson}. A more sophisticated method for direct creation of vortex
solitons may imprint the necessary phase pattern onto fundamental QDs by
vortical laser beams \cite{laser}.

Finally, it is relevant to estimate physical parameters of the modes
predicted in this work, with scaled norm $N$. Using the values of the
magnetic moment for $^{164}$Dy atoms and a typical transverse-confinement
length, $a_{\bot }\sim 0.5$ $\mathrm{\mu }$m \cite%
{Ramachandhran,1Li,2Li,Huang}, we conclude that the number of atoms in AVQDs
is $N_{\mathrm{atom}}\sim 10N$, and the scaled length unit corresponds to $%
\sim 0.1$ $\mathrm{\mu }$m.

\emph{Conclusion} We have constructed solutions for stable AVQDs
(anisotropic vortex quantum droplets) in the effectively two-dimensional
dipolar BEC. The anisotropy and stability are stipulated by the choice of
the polarization of atomic dipoles parallel to the system's plane and
perpendicular to the vortex' axis. The stability area of the AVQDs is
identified in the system's parameter space. Characteristic features of the
AVQDs, such as the peak density, chemical potential, effective area, aspect
ratio, and total angular momentum, are presented. Spinning AVQDs can stably
follow rotation of the polarizing magnetic field, provided that the rotation
is not too fast. Collisions between slow or fast moving AVQDs are elastic or
inelastic, respectively. In the latter case, the colliding AVQDs merge into
breathers. In particular, these may be bound states of the
vortex-antivortex-vortex type, which are also found as stable stationary
states. An efficient method for the creation of the AVQDs in the experiment
is to embed a phase kink into a zero-vorticity quantum droplet, letting it
spontaneously transform into an AVQD.

The present analysis can be extended further. First, it will be interesting
to apply initial torque to an elongated AVQD mode, and simulate ensuing
dynamics, which is expected to feature oscillations of the droplet's
orientation around the original elongated direction. Further, it may also be
relevant to simulate motion of a spinning AVQD, driven by the rotating
magnetic field, under the action of a kick. Another relevant possibility is
to construct AVQDs in binary dipolar BECs, cf. Refs. \cite%
{Gammatwocomponent,Ardila}. Finally, a challenging option is to seek for
stable AVQDs in the full 3D setting.

\begin{acknowledgments}
This work was supported by NNSFC (China) through Grants No. 12274077,
11874112, 11905032, by the Natural Science Foundation of Guangdong province
through Grant No. 2021A1515010214, and 2021A1515111015, the Key Research
Projects of General Colleges in Guangdong Province through grant No.
2019KZDXM001, the Research Fund of Guangdong-Hong Kong-Macao Joint
Laboratory for Intelligent Micro-Nano Optoelectronic Technology through
grant No.2020B1212030010. The work of B.A.M. is supported, in part, by the
Israel Science Foundation through grant No. 1695/22.
\end{acknowledgments}



\begin{thebibliography}{99}
\bibitem{Distance} The time should exceed the respective diffraction time,
at least, by an order of magnitude, be larger than $10\times 2R^{2}$, where $%
R$ is the radial size of the droplets. For example, in Figs. \ref{Example}%
(a) and \ref{v-av-v}(a), the radial sizes of the AVQD and the
vortex-antivortex-vortex self-bound state are $R\sim 30$ and $40$,
respectively, while the corresponding simulation times are $\simeq 18000$
and $32000$.
\end{thebibliography}
\end{document}